\documentclass[10pt,twocolumn]{article}

\usepackage[a4paper,margin=0.72in]{geometry}
\usepackage{graphicx}
\usepackage{amsmath}

\title{Simulative Anomaly Detection using 2D Tomography}
\author{
    Moti Ben-Harush\thanks{moti041@gmail.com, Elta Systems Ltd., Ashdod 7710202, Israel}
    \and
    Nimrod Teneh\thanks{nteneh@elta.co.il, Elta Systems Ltd., Ashdod 7710202, Israel}
    \and
    Gregory Lukovsky\thanks{glukovsky@elta.co.il, Elta Systems Ltd., Ashdod 7710202, Israel}
}
\date{}

\begin{document}

\maketitle

\begin{abstract}
We present a novel technique for predicting the imaging quality of anomalies
such as cancer cells located inside organic tissues. This technique is useful
for evaluating and designing RF tomography sensors.
\end{abstract}

\section{Introduction}

Research in the field of microwave ablation has increased in recent years due
to improved capabilities of antenna arrays. This treatment has the potential to
offer benefits both in detection and annihilation of tumors in living tissues
\cite{luyen2014microwave,bucci2016design}. The use of microwave radiation for
detection offers the ability to detect malignant tissues without harm to the
patient by injecting radioactive materials. Microwave ablation treatment has the
potential to annihilate tumors without the need to perform risky surgeries. The
fact that both exploit the observation that malignant tissues have different
dielectric properties than healthy tissues gives rise to the idea of performing
both procedures simultaneously.

In this paper we focus on malignant tissues imaging. We present a simplified
2D technique for predicting the imaging quality of anomalies, such as cancer
cells, located inside organic tissues. This technique is useful for evaluating
and designing RF tomography systems. In contrast to 3D tomography, 2D tomography
simulative approaches are significantly less time consuming; hence it is more
convenient for validation and robustness tests. The term 2D addresses anomalies
which obey translational invariance in one direction, for example an infinite
cylinder, which we refer to as ``2D anomalies'', see Fig.~\ref{fig:geometry}.
This simplification is justified due to high conductivity of the organic tissue.
The fact that the attenuation is proportional to the conductivity sets a limit
to the maximum length from which signal can be received.

\section{Analysis Description}

\begin{figure}[t]
    \centering
    \includegraphics[width=\linewidth]{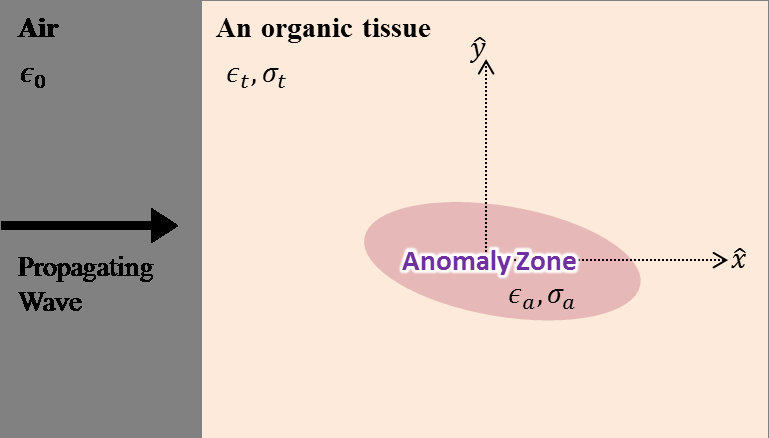}
    \caption{A description of the 2D geometry of an anomaly located inside the
    organic tissue. Each zone (air/tissue/anomaly) is separately characterized
    by relative permittivity and conductivity.}
    \label{fig:geometry}
\end{figure}

The geometry of our model is shown in Fig.~\ref{fig:geometry}. We consider a
planar wave propagating through space, impinging on the organic tissue which
contains a malignant tissue which defines the zone of anomaly. Each zone is
defined electromagnetically by frequency dependent relative permittivity and
conductivity.

The purpose of the presented analysis is to give an efficient way to estimate
the reduction in system performance in terms of image quality due to typical
inaccuracies inevitably existing in the reconstruction process. We believe that
the proposed analysis can provide an estimation for the robustness of the image
reconstruction procedure, which projects on the ability to empirically detect
anomalies in different types of tissues.

\subsection{Model}

Our model is based on 2D scattering equations for anomalies \cite{persico2014gpr},
which are formulated in matrix form and can be solved by using a variety of
matrix inversion methods \cite{persico2005role}. The inverse matrix is used to
form an image of the area of interest by using RF tomography. Each pixel in the
image gives an indication which is proportional to the relative dielectric
permittivity of the tissue at a given frequency.

One key ingredient of the procedure described above is the electric field in the
area of interest without any scatterers, which we define as the incident field,
\(E_{\mathrm{inc}}\) \cite{persico2005role}. This ingredient strongly affects
the quality of the reconstruction process of the image. The incident field is
prone to suffer from severe inaccuracies; thus it is important to examine its
influence over the reconstruction process.

The selection of operating frequency must take into account the attenuation of
propagating waves through the tissue. Therefore it is convenient to select a
frequency with a wavelength of the same length scale as the depth used for
imaging. Note that the effective resolution which sets the size of anomaly that
can be detected is also frequency dependent and should be taken into account.

\subsection{Simulation}

We used the WIPL-D 2D Solver, which is part of the WIPL-D software package
\cite{wipld2017}. The simulation time was of the order of several minutes, which
allows performance of a large number of robustness tests. The simulation
consisted of a planar wave propagating along the \(x\)-axis, and a set of
sensors located on a line parallel to the \(x\)-axis but shifted in some
distance \(d\). The sampling of the near-field distribution (NFD) of the
electric field was done at the locations marked by black asterisks. We chose the
location of the anomaly to be at the origin, hence the depth of the anomaly is
set by a shift of the location where the tissue begins. This can be seen in
Fig.~\ref{fig:error5}. The electric field measured by the sensors was provided
by a simulation of the NFD of the electric field inside the tissue.

\section{Results}

\begin{figure*}[t]
    \centering
    \begin{minipage}{0.48\textwidth}
        \centering
        \includegraphics[width=\linewidth]{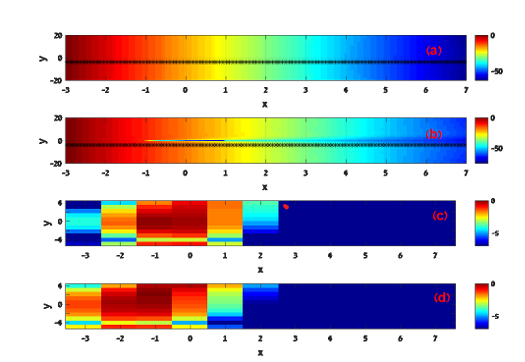}
        \caption{An analysis with an error of 5\% in the relative permittivity.
        The center of the anomaly is located at the origin. (a) NFD of the
        organic tissue without an anomaly in dB. (b) NFD of the organic tissue
        with an anomaly in dB. (c) RF tomography with accurate
        \(E_{\mathrm{inc}}\). (d) RF tomography with inaccurate
        \(E_{\mathrm{inc}}\).}
        \label{fig:error5}
    \end{minipage}
    \hfill
    \begin{minipage}{0.48\textwidth}
        \centering
        \includegraphics[width=\linewidth]{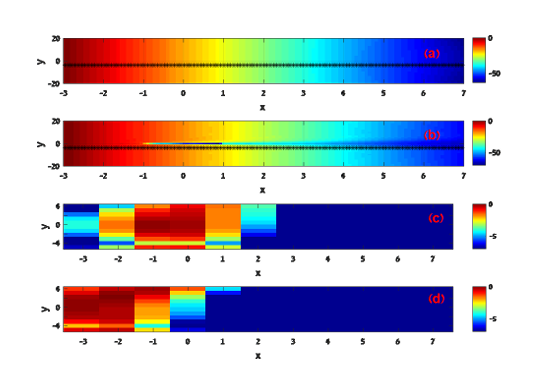}
        \caption{An analysis with an error of 15\% in the relative permittivity.
        The center of the anomaly is located at the origin. (a) NFD of the
        organic tissue without an anomaly in dB. (b) NFD of the organic tissue
        with an anomaly in dB. (c) RF tomography with accurate
        \(E_{\mathrm{inc}}\). (d) RF tomography with inaccurate
        \(E_{\mathrm{inc}}\).}
        \label{fig:error15}
    \end{minipage}
\end{figure*}

In Fig.~\ref{fig:error5}a and Fig.~\ref{fig:error5}b we present the NFD of the
electric field with and without an anomaly respectively. Note that the incident
field is easily given in the simulation; in the case of real measurement, the
measurement of the incident field is not trivial. An RF tomography image was
generated from the NFD results, see Fig.~\ref{fig:error5}c. As can be seen, the
tomography image detects the anomaly at the correct depth.

As mentioned above, a measurement of the incident field is easy with simulation,
but rather problematic in a real experiment. Hence we decided to test the
variability of our image reconstruction process to errors in the incident field.
In Fig.~\ref{fig:error5}d we present an RF tomography image, where we introduced
a 5\% error in the relative permittivity of the incident field. As can be seen,
the anomaly is still detected. Nevertheless, the image is slightly blurred, and
the depth of the anomaly is inaccurately estimated. We further increased the
error in the relative permittivity to 15\% and performed the same analysis. In
Fig.~\ref{fig:error15}d we present the result. As can be seen, the depth of the
anomaly is shifted to a non-realistic value. This sets a maximum level of
accepted error.

\section{Conclusions}

We presented an example for an analysis which provides an estimation for the
robustness of our imaging process for a given model, and electromagnetic
properties. Note that this analysis can be done for different parameters and a
wide variety of geometries.

An additional advantage of our approach is the fact that the signal processing
of the presented technique is based on scattering equations which restrict the
spectrum to a single frequency component. Thus it provides low sensitivity to
white noise \cite{persico2005role}. This fact is in line with the simplified
simulative approach which neglects this kind of noise.

\end{document}